# 20 years of network community detection

Santo Fortunato[1, 2] and M. E. J. Newman[3]

[1] Indiana University Network Science Institute (IUNI)

[2] Luddy School of Informatics, Computing and Engineering, Indiana University

[3] Department of Physics and Center for the Study of Complex Systems, University of Michigan

**A fundamental technical challenge in the analysis of network data is the automated discovery of communities — groups of nodes that are strongly connected or that share similar features or roles. In this commentary we review progress in the field over the last 20 years.**

Networks feature in many aspects of our lives, ranging from social networks of personal relationships, both on and off-line, biological networks of interactions between genes, metabolites and neurons, to technological networks such as the Internet, infrastructure networks, and transportation networks. The modern science of networks investigates the structure and function of networks like these, composed of nodes representing the elementary units of the system and the links or edges that connect them[1].

A prominent feature of networks is their community structure — the organization of nodes into groups, where nodes in the same group are strongly connected or share similar features or roles (Fig. 1). Algorithmic methods for the detection of communities in networks have found applications in a wide range of disciplines[2]. Early work on the problem goes back to the sociological literature of the 1980s[3], but the issue came to the wider attention of the physics community 20 years ago following the work of Girvan and Newman[4]. Here we review the advances of the intervening two decades in the detection of community structure in networks.

## Overview of the main approaches

Community detection is a rich and challenging problem, partly because it is not very well posed: what exactly do we mean by a community? In most cases, communities are defined as non-overlapping groups of nodes such that there are more edges within groups than between them, but this definition still leaves open many possibilities, and there are correspondingly many computational approaches.

The most common approaches are based on optimization. One assigns a score to each possible division of a network into communities such that 'good' divisions get high scores, then looks for the division with the highest score. There are a variety of ways to assign scores. The most popular approach makes use of the quality function known as modularity[5,6], which explicitly favours divisions with many edges within groups. The modularity is defined as the difference between the fraction of edges within groups and the expected fraction in a randomized version of the network. Formally, modularity can be written as a spin model, similar to an Ising or Potts model on the network, and a range of techniques from statistical physics brought to bear on its optimization and interpretation, including Markov chain Monte Carlo[7] and discrete optimization methods[6,8].

Another variant of the optimization approach, which has gained significant attention in recent years, is based on statistical inference. In this approach communities are not merely a feature of the network structure but a primary driver of it: nodes are connected precisely because of the groups they belong to, as might be the case for people with similar interests in a social network or web pages with similar topics. The placement of edges is represented using a probabilistic model such as the stochastic block model (SBM)[9], in which the probability that two nodes are connected depends solely on the communities they belong to. A 'good' community structure is one for which the model generates the observed network with high probability, so that the probability can be used as a score function for finding the best division. The calculation is complicated by the fact that the parameters describing the

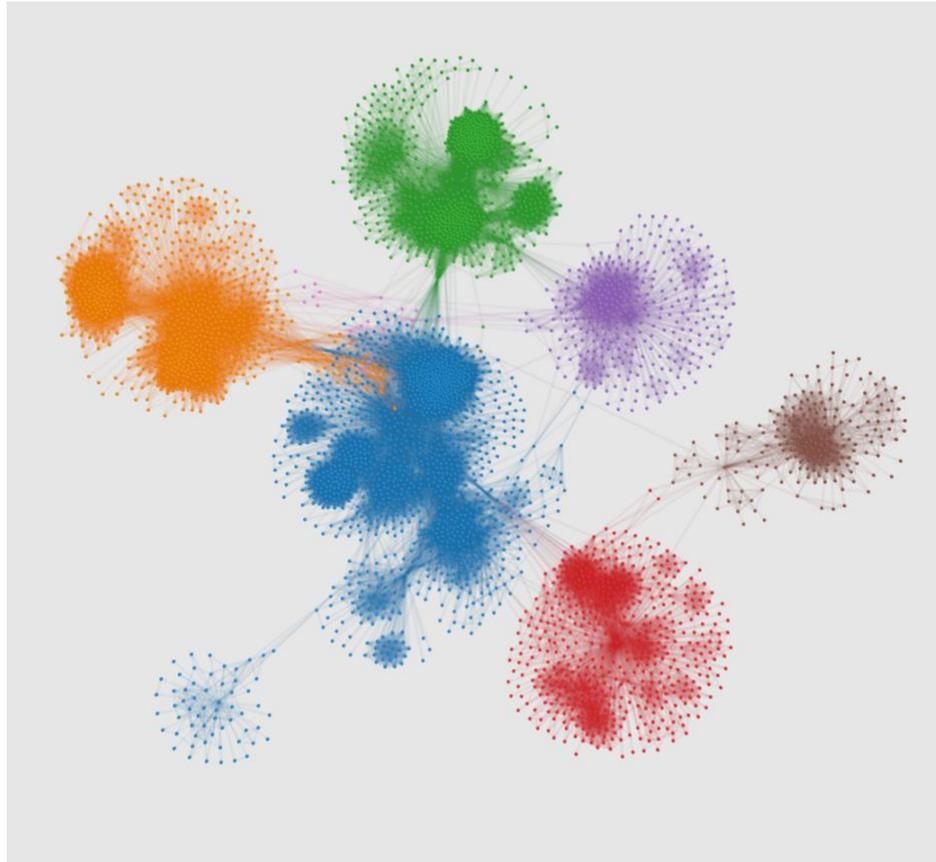

**Figure 1: Community structure of a social network**

Nodes are Facebook users and edges represent Facebook friendships. Communities, represented by different colours, were found using the InfoMap algorithm[11].

probabilities of the edges are usually unknown. To get around this, one can either optimize separately over these parameters or integrate them out. The former approach is more numerically efficient but the latter has the advantage of not requiring knowledge of the number of communities in advance. There also exists a microcanonical version of the SBM that fixes not the probabilities but the numbers of edges, yielding a different method of community detection based on minimum description lengths[10].

A third major class of community detection algorithms relies on links between community structure and dynamical processes taking place on networks, particularly random walks. As network communities are typically only sparsely connected to one another, a random walk on a network will tend to linger within communities and hence one can use the probability distribution of random walks for community detection. In the most popular method of this type, known as InfoMap[11], one considers the sequence of communities visited by a random walker — much like a road trip can be described by the sequence of regions visited — and a partition is considered good if the description of that sequence requires very little information, in the sense of Shannon entropy. A nice feature of this approach is that one does not have to actually perform any random walks to calculate the information: there is a closed-form expression for the entropy of an infinitely long random walk that one can use directly as a quality function for community detection.

## Global versus local

Although they work well, all of these methods for community detection depend on the structure of the entire network, which is not always ideal. Sometimes one might want to know the communities in only a small region and would like to avoid the computational burden of analysing the whole thing. More importantly the focus on complete networks can lead to problematic results in some situations. It seems implausible that the membership of a community should depend on the structure of portions of the network that are very far away — that the membership of a group of friends in a social network, for example, should be influenced by people in distant countries with whom they have no contact. And yet this is exactly what happens in some of the methods we have described.

Modularity maximization, for example, returns results that depend on the overall size of the network, giving rise to a so-called resolution limit that prevents the method from finding small communities in large networks[12]. Analogous limits also arise in other methods, such as those based on SBMs[10]. These problems can be mitigated by introducing parameters that control the resolution scale[13] or by constructing a nested hierarchy of divisions that allow one to probe smaller community scales[14]. But another option is to adopt a local approach to community detection in which individual communities are found without analysing the entire network. Local detection methods typically work by building a single community around a specified seed node, adding nodes greedily until a local optimum of some quality function is reached, an approach that can be highly computationally efficient[15,16].

## Benchmarks and performance tests

Given the large number of competing approaches for community detection, a critical question is how to choose among them. Performance is often measured in terms of an algorithm's success at recovering known communities planted in artificial benchmark networks. Such networks can be generated for instance using the SBM, a common approach in early works[4,17], but networks generated by SBMs differ substantially from their real-world counterparts, which are characterized by distinctive heterogeneous distributions of degree and community size[18]. To address this issue, more realistic variants of SBMs have been proposed, such as the degree-corrected SBM[19] and the LFR model[20], with the latter emerging as the standard benchmark in recent years. Algorithms' success at recovering the planted communities in these networks is typically estimated using information theoretic measures such as the normalized mutual information[17,21].

A number of large-scale studies have been carried out to compare the performance of the various community detection methods[17,22]. All of the approaches described above are found to have generally good performance on benchmark tests, though an interesting twist appears when the community structure in the test networks becomes weak. When the probability of edges is much higher within groups than between them, algorithms find the planted structures with ease. But as the probabilities become more similar community detection becomes harder, until communities vanish altogether when the probabilities are equal. Surprisingly, however, community detection actually fails before reaching this point: there is a region of non-vanishing size in the parameter space in which communities exist within the network but the algorithms fail to find them. It has been shown formally that there is a sharp phase transition point, known as the detectability threshold, below which all algorithms must fail to find the buried structure[23-25], a result that is instructive not only for community detection but also for network science more generally. It tells us that there are questions about networks that have well-defined answers and yet it is impossible to ever discover those answers.

An alternative approach for quantifying the performance of community detection algorithms is to measure their ability to recover known ground-truth communities in real networks. This has the advantage of testing performance in real-world settings, but the disadvantage that the ground-truth communities are rarely known exactly, so there is some ambiguity about what the correct answer is. In many cases, one uses some kind of node attributes or metadata as a proxy for the ground truth, but there is no guarantee that such metadata correspond exactly to communities, although there is empirical evidence that the two tend to be correlated[26]. Alternatively, one can exploit this correlation and create algorithms that infer communities by combining knowledge of metadata with network structure, and such algorithms, at least in some studies, appear to outperform those based on structure alone[27,28].

**Community overlap, hierarchy, embeddings**

So far we have focused on the classic problem of dividing a network into discrete, non-overlapping communities, but there are also a range of variants and extensions to other kinds of structure. For example, communities can subdivide into successively smaller communities through a hierarchy of levels that captures structure at scales from the entire network to the individual node[29]. Overlapping or mixed-membership communities in which the same node can belong to more than one group are common in social networks and some other networks[30,31]. Core-periphery structure describes networks that divide into groups of high and low density, with a dense core and a sparser periphery, or a larger nested set of groups with varying density[32,33]. All of these structure types can be detected using, for example, variants of modularity or statistical models.

A little further afield lie networks with latent space structure or stratification. These are networks in which the nodes possess (or can be assigned) numerical values that correlate with the placement of edges. An example is age stratification in social networks: people tend to be friends with others of about the same age, so edges fall between nodes with similar age values. Stratification can be thought of as a kind of community structure with continuous, scalar-valued community labels instead of discrete categorical ones. One can also have multiple numerical values on each node and think of them as coordinates that locate the nodes in a space, with edges that tend to fall between nearby nodes, an approach known as network embedding[34,35]. There exists a range of algorithms for finding embeddings of networks in low-dimensional spaces that can effectively infer node values even when they are not known in advance. Embedding a social network in a one-dimensional space, for example, could allow one to infer approximate ages for individuals even in the absence of any data about the individuals' identities.

Network embedding algorithms include classic approaches like spring embedding algorithms and Laplacian spectral embedding, which were originally developed with network visualization in mind, and statistical approaches more analogous to a continuous version of community detection[36,37]. Largely independent of these efforts, but arguably addressing similar questions, has been the development within computer science of methods for network representation learning, which attempt to assign each node in a network a score along one or more dimensions such that nodes that are similar in some sense will get similar scores. In practice, representation learning methods are rather different from the methods described in this Comment, often employing deep learning or neural network approaches. However, if the resulting scores are regarded as coordinates in a geometric space representation learning can give similar results to embedding methods. In addition to being of interest in their own right, embeddings also provide another route to the detection of standard discrete communities: one can define communities as groups of nodes that are close together in the space of the embedding and use classic data clustering methods to detect such groups.

**Outlook**

Community detection in networks is an area of vigorous ongoing research. The development of new algorithms continues to be of intense interest, with a particular focus on improving accuracy, providing formal guarantees of performance, and developing computationally efficient methods for application to very large data sets. Within mathematics and theoretical computer science there has been and will be extensive work on limits to algorithm performance and formal detectability bounds. Another area of interest is the development of new statistical models for network structure, both for the generation of benchmark networks and also for use in inference algorithms for communities and other more general forms of structure. And we envision further progress on measures, particularly based on information theoretic principles, for characterizing communities, comparing alternate divisions of a network, and clustering divisions into representative groups. Advances in representation learning will allow us to exploit multiple data types for community detection, not only the structure of the network, offering new powerful ways to solve the problem. Finally, there has been a dizzying array of applications of community detection methods, shedding light on the structure of networks in physics, biology, engineering, computer science, the social sciences, and beyond.

## Acknowledgements

S. F. gratefully acknowledges support by the Army Research Office under contract number W911NF-21-1-0194, by the Air Force Office of Scientific Research under award number FA9550-19-1-0391 and by the National Science Foundation under award number OISE-1927418. M. E. J. N. acknowledges support from the National Science Foundation under award number DMS–2005899.